\documentclass[preprint,12pt]{elsarticle}
\input{glyphtounicode}
\usepackage[T1]{fontenc}
\usepackage[utf8]{inputenc}
\usepackage{lmodern}
\DeclareUnicodeCharacter{2013}{--}
\DeclareUnicodeCharacter{2022}{\textbullet}
\usepackage{amssymb}
\usepackage{amsmath}
\usepackage{amsthm}
\usepackage{natbib}
\usepackage{float}
\usepackage{url}
\usepackage{hyperref}
\usepackage{booktabs}
\usepackage{algorithm}
\usepackage{algpseudocode}
\usepackage{makecell}

\graphicspath{{./figs/}}

\journal{Computer Languages, Systems \& Structures}

\begin{document}

\title{MSLL: A Runtime Multi-Stack Parsing Approach for Interactive Grammar Development\\
\large A Lightweight Extension of LL-Style Recursive Descent}

\author[1]{Qunhui Zhang}
\ead{will\_zhang@sjtu.edu.cn}

\address[1]{School of Software, Shanghai Jiao Tong University, Shanghai 200240, China}

\begin{abstract}
Few grammars are designed in a straight line. The usual rhythm is closer to trial and error: write a rule, parse a small input, discover that two alternatives fight over the same prefix, then either refactor the grammar or regenerate the parser and look again. Predictive parsers serve stable grammars very well; during this exploratory phase, however, they push work out of the editor and into an offline generation step. ANTLR-style LL(*) parsing is a good example: adaptive prediction and generated parser code deliver strong production performance, yet every grammar edit still has to pass through a tool-chain round trip before its effect can be inspected.
This paper describes MSLL, a small runtime extension of recursive-descent LL parsing. The mechanism is simple on purpose. When a FIRST/FIRST conflict appears, the parser keeps several live stacks at once, lets each one follow a different candidate production, and drops a stack the moment the token stream stops agreeing with it. Ambiguity is handled as a state-management problem at run time, not as something that must be engineered away before execution.
MSLL targets the period before a grammar settles. In our prototype sessions the grammar file changed more often than the input program did: a delimiter moved, a block rule got split in two, an object-literal alternative was left unresolved for one more test run. What matters in that setting is how quickly the parser can restart from an edited grammar and whether it can show which alternative survived. Research on incremental parsers, grammar workbenches, language-server integration, and error-message quality already treats edit-time behavior as a first-class concern; we apply the same requirement to grammar authoring itself.
The evaluation is mainly about locating the boundary of the approach. MSLL finishes the three 1k-token benchmarks in 48, 49, and 59 ms. On the two ambiguous 50k-token workloads it stays under one second (771 and 730 ms); the deterministic 50k-token stream costs 1.22 s, mostly because the prototype still performs full tokenization and tree bookkeeping. The nested-object stress case peaks at 85 MB of heap and 33 live stacks at 50k tokens, and the same case takes 5.395 s at 500k tokens. We read these numbers as a clear statement that MSLL is an edit-time parser, not a replacement for generated production parsers. A fourth case, a real-world JSON grammar over real tweet data, confirms the interactive range (6.6 ms at 1.2k tokens, about one second at 50k) and exposes a prototype-level super-linear cost at 500k tokens.
MSLL should therefore be seen as a development-time companion to existing LL-based parser generators: run and debug an evolving ANTLR-style grammar immediately, and hand the stabilized grammar to a generated parser once production speed becomes the dominant concern.
\end{abstract}
\maketitle
\section{Introduction}
Parsing is one of the few parts of a compiler that manages to be mathematically clean and intensely practical at the same time. A handwritten LL(1) parser for a small language is easy to read, easy to single-step, and cheap to run. The trouble starts while the language is still taking shape: alternatives overlap, left factoring keeps being postponed, and the designer wants to try a rule out before deciding whether it deserves to become permanent.
Classical LL(k) parsing adds fixed lookahead on top of LL(1), but the depth k must be picked in advance and larger values inflate the prediction structures quickly [1-2]. ANTLR's adaptive LL(*) parsing removes most of that burden with generated prediction machinery [3-4]. For production parsers this is exactly the right trade-off. For rapid grammar editing it is less comfortable, because the feedback loop still passes through parser generation, compilation, and sometimes an IDE refresh.
MSLL is built for the development phase rather than the deployment phase. It runs an ANTLR-style grammar directly and defers ambiguous decisions until the tokens themselves rule alternatives out. At every conflict point the active stack is forked; each fork keeps its own symbol stack, input position, and partial tree, and a fork that fails simply disappears. The programmer gets to watch the grammar behave without first reshaping it to satisfy a deterministic prediction table.
We deliberately stop short of GLR/GLL-level generality. GLR and GLL handle arbitrary context-free grammars and rely on shared representations to keep ambiguity compact [5-6]. MSLL keeps the recursive-descent execution model in plain view instead, retaining only the candidate top-down stacks that the current decision needs. The smaller design suits grammar-preview tools well, though it still exposes exponential corner cases whenever ambiguity persists across many tokens.
Once the grammar stabilizes, the supported MSLL dialect can be translated into an equivalent ANTLR .g4 grammar after a compatibility check and any required syntax adaptation [7]. The intended workflow has two stages: MSLL while the grammar is volatile, and a generated parser once the grammar is stable enough for production.
\section{Contributions}
This paper makes the following contributions:
\begin{itemize}
  \item We introduce MSLL, a runtime extension of recursive-descent LL parsing. At a FIRST/FIRST conflict it keeps one stack per candidate production and records where each stack later dies, so a grammar author can study an execution trace before deciding whether to factor or rewrite the rule. Unlike GLL-style generalized parsing, MSLL builds no shared packed parse forest, inserts no generation step between grammar and execution, and treats the conflict trace itself as the primary output.
  \item We implement the approach for a limited ANTLR-style grammar dialect while keeping the runtime machinery small: grammar loading, non-deterministic prediction lookup, stack management, and parse-tree bookkeeping.
  \item We specify MSLL's operational behavior and discuss soundness, completeness relative to the explored top-down derivations, and the practical boundary that stack growth imposes.
  \item We evaluate a Java prototype on three DSL-style workloads, with inputs from 1k to 500k tokens, reporting wall-clock time, heap footprint, cumulative stack allocations, and peak live-stack counts against rewritten LL(1) and generated ANTLR baselines, plus a fourth case that runs a real-world JSON grammar over real inputs.
  \item We relate the algorithm to current language-tooling practice, including parser generator evolution, incremental parsing systems, error-tolerant parsers, parallel parsing engines, and textual DSL workbenches [7-28].
\end{itemize}
\section{Related Work}
Parsing research has accumulated several traditions: deterministic predictive parsing, generalized parsing, parser combinators, parsing expression grammars, and language workbenches. MSLL takes the readable control flow of recursive descent and treats ambiguous prediction as a runtime scheduling problem. Below we position that choice against both the classic parsing literature and recent tool-oriented work.
\subsection{Predictive LL Parsing}
Classical LL(1) parsing relies on a single-token lookahead and requires
grammars to be fully disambiguated before execution. In a compiler textbook or a settled production grammar that restriction is perfectly reasonable. It is far less convenient while a DSL is still moving and the author would rather watch a rule behave before refactoring the syntax around it.
Handwritten recursive-descent parsers stay popular for small languages, mostly because the control flow mirrors the grammar; the same rigidity, though, becomes expensive during early-stage grammar design.
LL(k) parsing generalizes LL(1) by inspecting k tokens, but k still has to be fixed before execution and larger prediction structures come with it [1-2]. ANTLR's adaptive LL(*) and ALL(*) machinery shifts that work into generated prediction code and runtime analysis [3-4]. Current ANTLR releases also support a broad set of target languages, one reason ANTLR remains the natural deployment target after a grammar stabilizes [7].
Recent parser-generator work suggests that engineering convenience is still an open research issue. Paguroidea rethinks the lexer/parser boundary and keeps semantic actions transparent to the user [14]; LarkAG adds attribute-grammar support to a widely used grammar tool [17]. Neither system makes the runtime choice MSLL makes, but both confirm that parser tools continue to be shaped around maintainability rather than asymptotic throughput alone.
MSLL moves the friction point into the runtime instead. The goal is not to displace optimized generated parsers. It is to give grammar authors a temporary execution mode in which overlapping alternatives can be inspected directly and resolved later, once the language design has firmed up.
\subsection{Dynamic and Adaptive LL Parsing}
A different line of work attacks ambiguity from the verification side. Palmkvist et al. study when ambiguity can be resolved statically for particular grammar forms [29], and Morpheus verifies safety properties of data-dependent parser-combinator programs [30]. MSLL is less proof-oriented; it maintains a concrete set of runtime alternatives and reports how they disappear over a grammar edit-run cycle.
The path MSLL takes is operational. Rather than computing a sufficient lookahead bound or establishing a grammar class up front, it keeps every currently viable top-down path alive and removes a path when a token contradicts it. During debugging, this explicit behavior tends to be easier for developers to reason about than a generated prediction artifact.
\subsection{Generalized Parsing Algorithms}
GLR and GLL sit at the general end of the design space. GLR keeps several LR actions alive and compacts ambiguity; GLL brings graph-structured stacks and descriptor scheduling into a recursive-descent setting [5-6]. Happy-GLL extends the line with reusable top-down parsers for parameterized nonterminals [31]. MSLL borrows the keep-alternatives-alive idea but skips full parse-forest construction, since the target is quick grammar inspection rather than exhaustive ambiguity enumeration.
That generality has a price: machinery a lightweight DSL editor may simply not need. MSLL maintains no shared packed parse forest and makes no attempt to enumerate every ambiguity for later semantic resolution. Three concrete choices separate it from a GLL-style engine. Stacks are plain copies rather than nodes in a graph-structured stack, so the runtime remains a readable recursive-descent loop. The parser is interpreted straight from the grammar file, with no generation step in between. And the artifact handed back to the author is the trace of fork and prune events, not a parse forest. In exchange, MSLL gives up the worst-case efficiency guarantees that GSS sharing provides, a trade we consider acceptable for an edit-time tool.
\subsection{Parsing Expression Grammars and Combinators}
Parsing Expression Grammars settle ambiguity through ordered choice [32]. PEG and LPeg-style systems appeal because their execution model is predictable, and recent LPeg work shows that careful PEG design can yield efficient practical parsers [33]. Adaptable PEG libraries push the tradition further toward dynamic grammar modification [34].
Parser combinators give compositional grammar definitions, and verified or effect-aware combinator systems can now back stronger safety claims [30]. As a programming model MSLL is less expressive, but it stays much closer to the grammar-file workflow that ANTLR users already have.
What really differs is the treatment of failed alternatives. Under prioritized choice, a later PEG alternative can become invisible once an earlier one succeeds. In MSLL the competing alternatives stay visible until the input itself rejects them, which helps when an author is trying to work out why two productions overlap in the first place.
\subsection{Incremental and Interactive Parsing Tools}
Editor parsers make a related trade-off, but from the source-edit side. Tree-sitter keeps parse trees incrementally updatable [8]; Lezer favors compact trees, fragment reuse, and error-insensitive reparsing [9]; Langium wires grammars into language-server services for textual DSLs [10, 22]. All of these optimize edits to the program being parsed. MSLL moves the optimization point to edits of the grammar itself.
The distinction is practical. When only the source file changes, Tree-sitter- or Lezer-style reuse is the more direct answer. When the rule under test changes, a parse tree cached for the old grammar helps much less. MSLL's advantage is that the modified grammar runs immediately, so the author can look at the next conflict without ever producing new parser source code.
Language-workbench research reports the same pressure. Studies of Xtext-based DSLs find that grammar definitions, metamodels, and example instances tend to evolve together, often in small steps that a full regeneration cycle handles poorly [13, 15, 19-21]. Those small grammar steps, rather than the final generated parser, are what MSLL is aimed at.
\subsection{Correctness and Verification}
Correctness work for parsing stretches from classic grammar-class proofs to verified parser generators and verified parser combinators [29-30]. MSLL has no mechanized proof yet. What this paper offers is an operational argument: every surviving stack corresponds to a valid partial top-down derivation, and pruning removes only the paths the input has contradicted.
\subsection{Summary}
MSLL therefore occupies a practical middle ground: lighter than full generalized parsing, more exploratory than static LL parser generation, and more transparent than prioritized-choice execution when the job is debugging ambiguous grammar alternatives.
\section{MSLL Algorithm}
\subsection{Overview}
The Multi-Stack LL (MSLL) algorithm extends ordinary recursive-descent LL control flow with a set of active parsing stacks. Each stack records what one top-down derivation still expects to see. While prediction stays deterministic, MSLL behaves exactly like an LL(1) parser; once more than one production can start with the same lookahead token, the stack is forked and the competing productions run side by side.
This fork-and-prune behavior at run time spares the author immediate grammar rewriting and parser regeneration. It pays off most when the grammar is a design artifact under active editing rather than a stable artifact ready for production generation.
\subsection{Parsing Procedure}
Each parsing stack stores three pieces of state:
\begin{itemize}
  \item a work list of grammar symbols that this derivation still expects to match;
  \item the current token index in the input stream;
  \item the partial parse tree fragment produced so far.
\end{itemize}
The parser repeatedly advances all active stacks:
\begin{itemize}
  \item Terminal case: with a terminal t on top, the parser compares t against T[i]. A match advances i and removes t; a mismatch marks the stack as failed.
  \item Expansion case: with a non-terminal A on top, the parser asks the prediction table for every production of A whose FIRST information admits the current lookahead token.
  \item Deterministic expansion: a lookup that yields a single candidate rewrites the stack in place, with no branch object allocated.
  \item Multiple productions: when several productions apply, the current stack is copied once per production, each copy following one alternative.
  \item Failed case: an empty prediction set or a terminal mismatch removes the affected stack and nothing else; the remaining stacks continue from their own input positions.
  \item Pruning: a stack that reaches a contradiction is removed at once, so later steps only ever touch viable candidates.
  \item Completion: the parse succeeds when exactly one stack has consumed the whole input and emptied its symbol list. Anything else is reported as failure or as residual ambiguity.
\end{itemize}
\subsection{Formal Specification}
Let:
\begin{itemize}
  \item G = (N, $\Sigma$, P, S0) is a context-free grammar, where S0 is the start symbol;
  \item T = t0 t1 ... t(n-1) is the input token stream;
  \item Active is the current set of runtime stacks. Each stack s in Active maintains:
  \item a working symbol stack $\alpha$ over terminals and non-terminals;
  \item an input index i in the range 0 ... n;
  \item a partial parse tree $\tau$ and trace metadata.
\end{itemize}
The parser proceeds as follows:
\paragraph*{Terminal match}
\begin{itemize}
  \item If top($\alpha$) is terminal t and i < n and t = T[i], consume t, increment i, pop $\alpha$, and update $\tau$.
  \item If top($\alpha$) is terminal t and either i = n or t != T[i], terminate that stack.
\end{itemize}
Non-terminal expansion: let top($\alpha$) = A. Let Pred(A, T[i]) = \{A -> $\beta_{1}$, ..., A -> $\beta_{k}$\} be the prediction entry admitted by FIRST/FOLLOW information for the current lookahead.
\begin{itemize}
  \item If k = 0, terminate the stack.
  \item If k = 1, replace A with $\beta_{1}$ in place.
  \item If k > 1, duplicate the stack into k continuations and replace A with a distinct $\beta_{j}$ in each continuation.
\end{itemize}
\paragraph*{Termination}
Let Active\_final be the set of stacks with empty symbol lists and fully consumed input (i = $|T|$).
\begin{itemize}
  \item Success: if |Active\_final| = 1, return its parse tree.
  \item Failure or residual ambiguity: if |Active\_final| = 0 or |Active\_final| > 1, report the corresponding error or ambiguity trace.
\end{itemize}
\subsection{Correctness of MSLL}
The correctness argument is scoped to the derivations that active stacks represent and to the prediction relation the prototype implements. Cloning a stack must correspond to choosing one grammar production, and pruning may happen only after the next terminal or prediction entry contradicts the derivation. Under those conditions, duplicating a stack changes scheduling and nothing else; the grammar's language is untouched.
Assumptions. Let:
\begin{itemize}
  \item G = (N, $\Sigma$, P, S0) is a context-free grammar in the supported top-down subset;
  \item w is the input token sequence;
  \item L(G) denotes the language generated by G;
  \item every live runtime stack represents one partial leftmost top-down derivation.
\end{itemize}
Static LL(k) parsers depend on a fixed prediction depth, whereas ANTLR 4 performs adaptive LL(*) prediction at runtime by simulating its ATN and caching explored decisions as DFAs [1, 4]. MSLL instead carries alternative derivations around as runtime values, so the correctness question becomes whether those values faithfully represent possible derivations and whether pruning removes only the impossible ones.
Soundness. Claim. If MSLL returns a parse tree for input u, then u is in L(G).
Argument. The initial stack contains the start symbol and token index 0. Each later stack is obtained by a legal terminal match or by replacing a non-terminal with the right-hand side of one of its productions. A stack can be returned only after it consumes the entire token stream and has no pending grammar symbols. Its tree is therefore assembled from legal grammar steps, so the accepted input belongs to L(G).
Relative completeness. Claim. If there exists a valid top-down derivation of u that is admitted by the prototype's prediction relation, then MSLL retains at least one corresponding active stack until success.
Argument. At each ambiguous non-terminal, MSLL creates one continuation for every admitted production. A stack is pruned only after its next required terminal or prediction entry becomes incompatible with the input. Therefore, any admitted derivation that remains compatible with the consumed input has a corresponding stack until it either succeeds or is contradicted by the input.
Discussion. One virtue of MSLL is that the cost of unresolved ambiguity stays visible. When many alternatives remain viable across many tokens the stack count grows; when conflicts are local, pruning keeps the active set small. The same observation explains both why MSLL helps in interactive debugging and where its limits lie under pathological ambiguity.
\subsection{Formal Pseudocode of MSLL}
Algorithm 1 summarizes the runtime loop used by MSLL.
\begin{algorithm}[H]
\caption{MSLL Parsing Algorithm}
\label{alg:msll}
\begin{algorithmic}[1]
\State Initialize active\_stacks $\leftarrow$ \{([S], 0, empty tree)\}
\State while active\_stacks contains an unfinished state do
\State for each ($\alpha$, i, $\tau$) in active\_stacks do
\State if top($\alpha$) = a is terminal then
\State if i < $|T|$ and a = T[i], pop a; i $\leftarrow$ i + 1; update $\tau$; else terminate; continue
\State P $\leftarrow$ Pred(top($\alpha$), lookahead(T, i))
\State if $|P|$ = 0 then terminate the state
\State else if $|P|$ = 1 then apply P1 in place
\State else clone once per Pj and apply Pj to clone j
\State end if
\State end for
\State end while
\State return the unique complete state or report failure/ambiguity
\end{algorithmic}
\end{algorithm}
\section{System Design and Architecture}
The runtime architecture is intentionally small. Grammar loading, tokenization, prediction lookup, active-stack management, and parse-tree assembly live in separate components, which makes it straightforward to inspect a conflict, trace why a particular stack was pruned, and reuse the same grammar structure when preparing an equivalent grammar for a generated parser.
\subsection{Key Components}
\begin{itemize}
  \item Lexer: scans the source string once, emitting token kinds and lexemes and keeping offsets so conflict traces can point back into the input.
  \item Parser: drives the top-down expansion loop over the set of active stacks.
  \item Prediction Table: holds one or more candidate productions per non-terminal/lookahead pair.
  \item Stack Manager: allocates, clones, schedules, and prunes the parsing stacks.
\end{itemize}
\subsection{Processing Pipeline}
\begin{itemize}
  \item Tokenization: a single lexer pass returns token kind, lexeme, and source offset to the runtime parser.
  \item Parsing loop: active stacks are scheduled in a fixed order, so two runs over the same grammar yield the same trace. Each stack carries its own token index and tree frontier, which means the scheduler could later be parallelized without changing the algorithm.
\end{itemize}
Parallel execution is thus an implementation opportunity rather than an algorithmic requirement. Both the earlier parallel parsing literature and current multi-core engines for regular-text recognition and parsing [24-25] suggest that high-ambiguity workloads would benefit from a coordinated parallel stack scheduler.
\begin{itemize}
  \item Initial stack: the stack manager begins with the grammar start symbol, token index 0, and an empty tree frontier.
  \item Prediction lookup: the parser collects every production that can begin with the current lookahead token.
  \item Stack forking: a lookup that returns several productions makes the stack manager clone the current context.
  \item Stack cloning: clones share the immutable grammar metadata while carrying their own symbol lists, input positions, and tree fragments.
  \item Failure and pruning: terminal mismatches and empty predictions go into the trace, and the failed stack is dropped.
\end{itemize}
The architecture is built around direct execution of evolving grammars. Conflicts are not hidden inside generated code; they surface through active-stack traces and pruning events (Figure 1).

\begin{figure}[H]
  \centering
  \includegraphics[width=\linewidth]{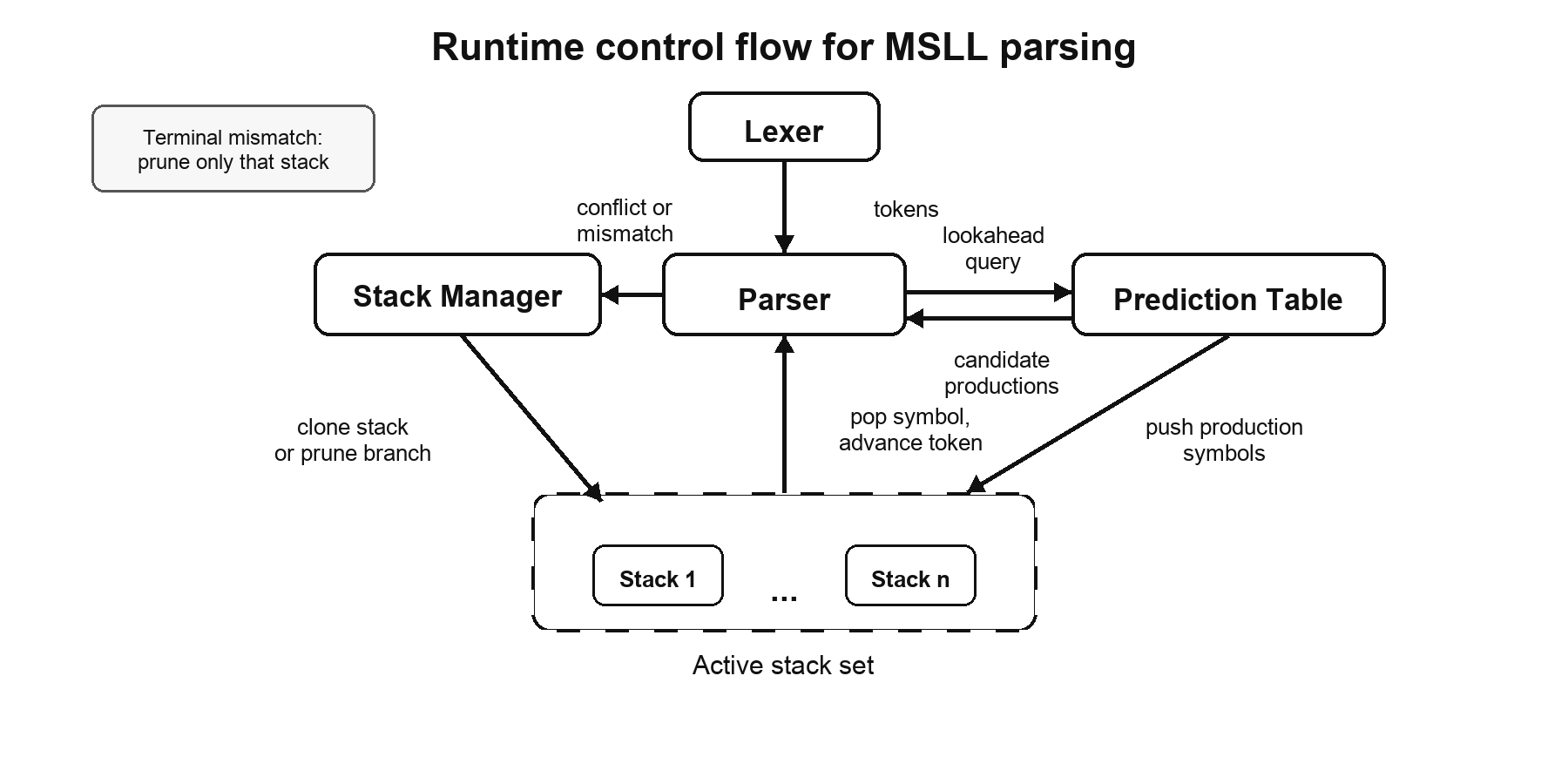}
  \caption{MSLL runtime architecture and active-stack data flow}
  \label{fig:msll-f1}
\end{figure}
\section{Implementation}
\subsection{Grammar Compatibility}
We implemented MSLL as a lightweight Java prototype, mainly to test whether multi-stack prediction can be embedded without rewriting an entire parser framework. The prototype reads grammars written in an ANTLR-style dialect, with parser and lexer rules in separate files and lexer rules expressed as regular expressions; it builds a rule table, computes FIRST/FOLLOW-based prediction entries for the supported constructs, and then executes the grammar directly. The prototype and the evaluation grammars are publicly available at https://github.com/WillCaptain/msll.
Grammar support currently covers the EBNF constructs the evaluation grammars need: grouping, repetition, alternatives, and ordinary terminal and non-terminal references. Direct left recursion and full ANTLR semantic predicates are outside the current prototype.
This compatibility matters for the proposed workflow, and not only for parsing: .g4 grammars are reused across a wider tool ecosystem, including grammar-based test generation [28]. An author can debug with MSLL while the language is changing, then translate the supported grammar into an equivalent ANTLR .g4 grammar when the syntax is stable [7]. This is structural compatibility, not direct file portability: the separate parser/lexer files, regular-expression lexer syntax, and unsupported ANTLR features may require adaptation.
\subsection{Parsing Pipeline}
The MSLL engine is organized as follows:
\begin{itemize}
  \item Grammar Loader: turns grammar rules into internal rule objects, normalizing the EBNF constructs the evaluation grammars use while preserving alternative order for trace output.
  \item Prediction Table Builder: computes FIRST sets and, where LL(1) would report a conflict, keeps multiple productions in the entry.
  \item Runtime Stack Engine: holds the active frontier of the parse. Each stack record contains:
  \item a working symbol list;
  \item an input pointer;
  \item a parse-tree fragment and trace metadata;
  \item Ambiguity Resolver: when a prediction entry holds several candidates, the first one continues on the current stack, the rest are cloned, and the input position where each branch later fails is stored.
\end{itemize}
A successful parse returns an annotated AST together with a compact trace of conflict decisions. A failed parse reports the point at which all stacks were pruned, which is exactly the information an author debugging an overlapping rule needs.
\subsection{Parser Initialization}
The API follows the builder style common in Java tooling libraries: supply parser and lexer grammar files, create a parser object, then call parse on a source snippet:
\begin{verbatim}
MyParserBuilder builder = new MyParserBuilder("testParser.gm", "testLexer.gm");
\end{verbatim}
\begin{verbatim}
String source = "... source code ...";
\end{verbatim}
\begin{verbatim}
MyParser parser = builder.createParser(source);
\end{verbatim}
\begin{verbatim}
ParserTree tree = parser.parse();
\end{verbatim}
Grammar loading and prediction-table construction happen at parser creation time, and no Java parser source is ever generated. After a rule edit, the prototype reloads the grammar file and rebuilds only the runtime tables the next parse needs.
\subsection{Optimizations}
The prototype uses three simple optimizations:
\begin{itemize}
  \item Greedy-first evaluation: the first compatible production stays on the original stack, which avoids needless allocation across the common deterministic regions.
  \item Lazy pruning: an invalid path is discarded at the earliest mismatch instead of after a full parse attempt.
  \item Prefix sharing for parse trees: alternatives that have consumed the same prefix point to immutable prefix nodes, so a clone usually copies stack metadata and a pointer instead of an entire subtree.
\end{itemize}
Together these measures let MSLL behave like ordinary LL parsing on unambiguous fragments while still preserving competing alternatives where the ambiguity is real.
\subsection{Example Demonstration}
Figure 2 shows MSLL forking two brace-delimited alternatives, then pruning the branch whose next expected terminal fails to match the input.

\begin{figure}[H]
  \centering
  \includegraphics[width=\linewidth]{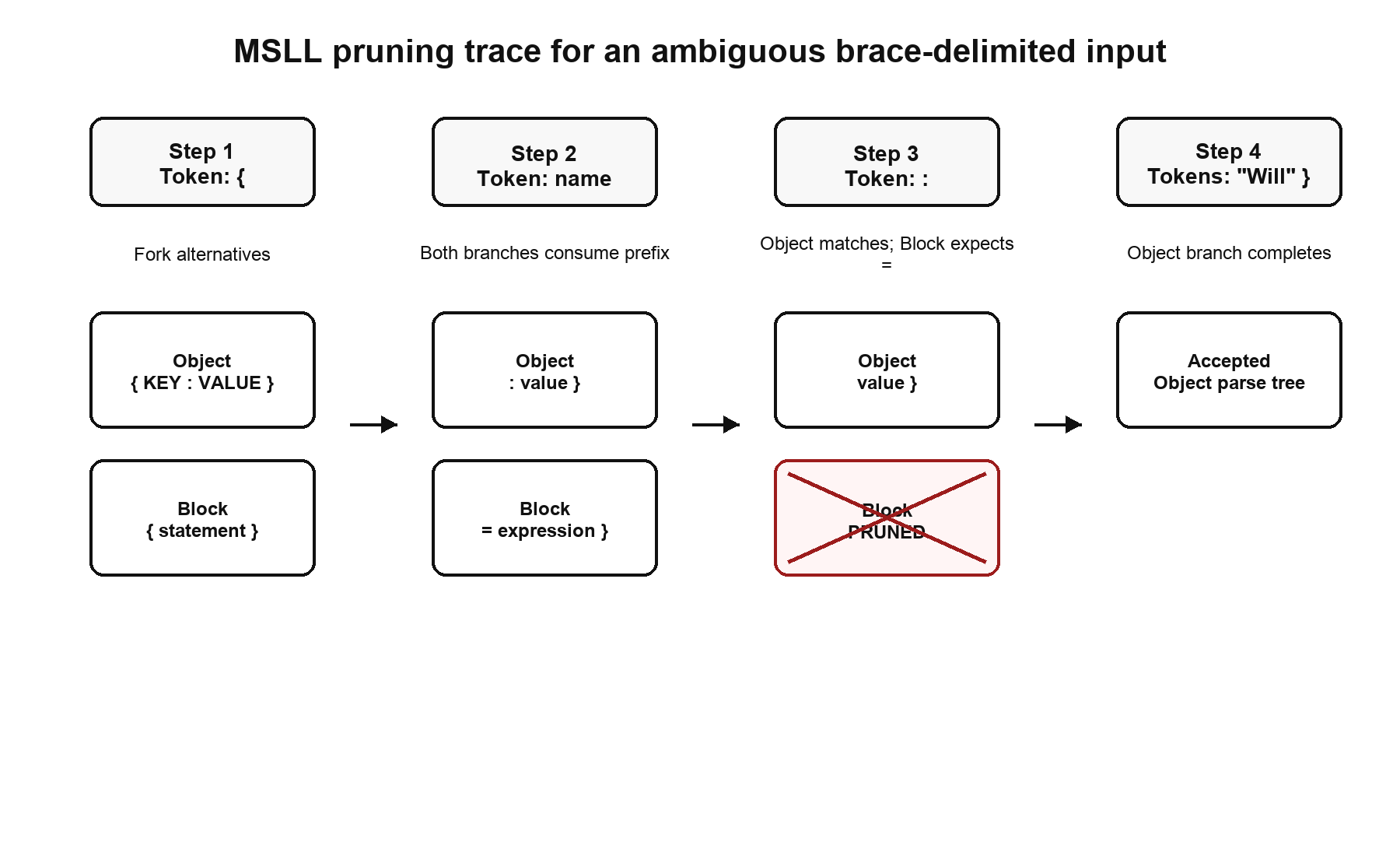}
  \caption{MSLL pruning trace for an ambiguous brace-delimited input}
  \label{fig:msll-f2}
\end{figure}
\section{Evaluation}
The evaluation separates three questions: correctness, runtime overhead, and edit-loop cost. We first compare returned trees against expected structures on controlled grammars, then record time, memory, and stack counts as ambiguity grows, and finally measure the latency a grammar author sees after a small rule edit. The point throughout is to locate the useful operating range of runtime ambiguity handling.
\subsection{Evaluation Goals}
We aim to answer the following research questions:
\begin{itemize}
  \item RQ1: Correctness. Does MSLL return the expected parse tree on grammars with controlled ambiguity?
  \item RQ2: Performance. As token counts grow from 1k to 500k, how much time and heap overhead does MSLL add over LL(1) and ANTLR?
  \item RQ3: Stack growth. How fast do cumulative and peak active-stack counts rise on deeply nested ambiguous inputs?
\end{itemize}
\subsection{Experimental Setup}
All measurements ran on a dedicated machine to limit interference from background tasks, and every parser configuration was warmed up before timing so that one-time JVM effects would not dominate. For each input size, the same grammar and token stream were fed to all parsers whenever the baseline could accept the grammar. The configuration was held fixed across all parsers; absolute times would of course shift on a newer JVM or faster hardware, but since every comparison in the tables keeps the runtime constant, the relative trends are what the numbers are meant to support. Case 4, added later on a different machine, states its own configuration in Section 7.6.
For steady-state timing, the ANTLR parsers were generated once before measurement and reused across timed runs, while MSLL loaded an equivalent runtime-dialect grammar at runtime. Parse throughput is therefore reported separately from the later edit-loop experiment, where code generation and compilation latency become part of what the developer actually feels.
\begin{itemize}
  \item Hardware: Intel i7 (3.1GHz), 16GB RAM
  \item Environment: Java 8, OpenJDK, GC enabled
  \item Baselines:
  \item – LL(1): custom handwritten recursive-descent parser (grammar rewritten to remove FIRST/FIRST conflicts)
  \item – ANTLR 4.13.0: code-generated LL(*) parser
  \item MSLL configuration: pruning enabled; greedy-first scheduling enabled
\end{itemize}
Every test uses a .g4-style grammar. ANTLR consumes it through the normal generation path; MSLL loads it directly and builds its non-deterministic prediction table at runtime.
\subsection{Case 1: Multi-Layer Code Blocks and JSON Structures}
The first workload models nested blocks and JSON-like object literals, a shape common in configuration DSLs and transformation scripts:
let me = \{ let age = 40; \}
\begin{verbatim}
let profile = {{{{{ age: age, name: { first: "Will" } }}}}};
\end{verbatim}
me.make\_friend(\{ name: "Alice", age: 30 \});
The grammar contains overlapping productions for brace-delimited constructs:
\begin{verbatim}
Block ::= '{' statement+ '}'
\end{verbatim}
\begin{verbatim}
Object ::= '{' (KEY ':' VALUE ',')* KEY ':' VALUE '}'
\end{verbatim}
Test Result:
\begin{table}[H]
\centering
\caption{Parsing results for multi-layer code blocks and JSON structures}
\label{tab:msll-table1}
\resizebox{\linewidth}{!}{
\begin{tabular}{lcccc}
\hline
Token Size(k) & Parser & Memory(MB) & Time(ms) & Stack Count total (max simultaneously) \\
\hline
1k & MSLL & 2–7 & 48 & 220(5) \\
1k & LL(1) & 2–7 & 35 & 1 \\
1k & ANTLR & 2–7 & 40 & 0 \\
50k & MSLL & 116 & 771 & 7301(5) \\
50k & LL(1) & 38 & 663 & 1 \\
50k & ANTLR & 83 & 713 & 0 \\
150k & MSLL & 121 & 2020 & 21901(5) \\
150k & LL(1) & 42 & 1955 & 1 \\
150k & ANTLR & 91 & 1896 & 0 \\
300k & MSLL & 128 & 3760 & 43801(5) \\
300k & LL(1) & 44 & 3440 & 1 \\
300k & ANTLR & 96 & 3560 & 0 \\
500k & MSLL & 143 & 5906 & 73001(5) \\
500k & LL(1) & 49 & 5320 & 1 \\
500k & ANTLR & 118 & 5430 & 0 \\
\hline
\end{tabular}
}
\end{table}
As Table 1 shows, MSLL pays a moderate price under nested ambiguity but stays inside the intended interactive range. The active-stack count remains low because the conflicting alternatives get pruned within a few tokens.
\subsection{Case 2: Pathological Ambiguity (Repeated Deep Nesting)}
The second workload repeats a nested object shape with heavier ambiguity:
\begin{verbatim}
me.make_friend({
  name: { first: "Will", friends: [{ name: "Alice", friends: [] }] }
});
\end{verbatim}
This case keeps more alternatives alive, peaking at 33 simultaneous stacks. As a stress test it is useful precisely because it exposes the duplication cost a grammar author would have to control before any production deployment.
\begin{table}[H]
\centering
\caption{Parsing results for repeated deep nesting}
\label{tab:msll-table2}
\resizebox{\linewidth}{!}{
\begin{tabular}{lcccc}
\hline
Token Size(k) & Parser & Memory(MB) & Time(ms) & Stack Count total (max simultaneously) \\
\hline
1k & MSLL & 2–7 & 49 & 541(33) \\
1k & LL(1) & 2–7 & 36 & 1 \\
1k & ANTLR & 2–7 & 40 & 0 \\
50k & MSLL & 85 & 730 & 18001(33) \\
50k & LL(1) & 60 & 570 & 1 \\
50k & ANTLR & 82 & 653 & 0 \\
500k & MSLL & 169 & 5395 & 180001(33) \\
500k & LL(1) & 125 & 4150 & 1 \\
500k & ANTLR & 146 & 4680 & 0 \\
\hline
\end{tabular}
}
\end{table}
\subsection{Case 3: Flat Deterministic Grammar}
The third workload repeats a flat deterministic assignment:
\begin{verbatim}
var  a  =  100;
\end{verbatim}
As expected, MSLL tracks LL(1) closely whenever prediction is deterministic and no stack fork is needed (Table 3).
\begin{table}[H]
\centering
\caption{Parsing results for a flat deterministic grammar}
\label{tab:msll-table3}
\resizebox{\linewidth}{!}{
\begin{tabular}{lcccc}
\hline
Token Size(k) & Parser & Memory(MB) & Time(ms) & Stack Count total (max simultaneously) \\
\hline
1k & MSLL & 2–7 & 59 & 1 \\
1k & LL(1) & 2–7 & 55 & 1 \\
1k & ANTLR & 2–7 & 57 & 0 \\
10k & MSLL & 63 & 297 & 1 \\
10k & LL(1) & 52 & 254 & 1 \\
10k & ANTLR & 58 & 273 & 0 \\
50k & MSLL & 124 & 1223 & 1 \\
50k & LL(1) & 104 & 1093 & 1 \\
50k & ANTLR & 112 & 1125 & 0 \\
\hline
\end{tabular}
}
\end{table}
\subsection{Case 4: A Real-World JSON Grammar on Real Inputs}
The three controlled workloads isolate ambiguity by construction, so we added a fourth case built entirely from public material. The grammar is the JSON grammar from the grammars-v4 collection, transcribed into the prototype's ANTLR-style grammar dialect: the rule structure is unchanged, including the two FIRST/FIRST conflicts that the object and array rules place on the opening brace and bracket, while the lexer rules are expressed as regular expressions and the prototype's automatic-epsilon option is enabled. The inputs are real tweets from the nativejson-benchmark twitter.json corpus, sliced and replicated to 1,248, 50,124, and 500,091 tokens (counts taken from the ANTLR lexer); no field is generated.
This case ran on a different configuration from Section 7.2: an Apple M1 Pro with OpenJDK 21. Comparisons are therefore internal to Table 4, and absolute values must not be read against Tables 1-3. The LL(1) baseline is omitted because the real grammar was not rewritten into a conflict-free form; MSLL is compared against the generated ANTLR parser only. Warm-up and timing follow the protocol of Section 7.2, and each reported time is the median of ten timed runs.
Table 4 shows two things at once. On editing-scale inputs the prototype behaves as intended: 6.6 ms at 1.2k tokens and about one second at 50k, with a fork raised roughly once every twelve tokens, matching the frequency of braces and brackets in tweet JSON, and never more than 4 stacks alive at once because every fork is resolved by the very next token. At 500k tokens, however, the prototype needs 85 s where ANTLR needs 172 ms. The peak stack count is still 4 and total forks still grow linearly, so stack duplication is not what hurts; the super-linear cost points to per-fork bookkeeping that grows with the consumed prefix. We state this plainly: it is an implementation limitation rather than a property of the multi-stack scheme, it bounds the input sizes for which the current prototype stays interactive on real data, and it is consistent with the position that stabilized grammars and large inputs belong to a generated parser. Section 12 lists profiling and removing this overhead as immediate future work.
\begin{table}[H]
\centering
\caption{Parsing results for a real-world JSON grammar on real inputs}
\label{tab:msll-table4}
\resizebox{\linewidth}{!}{
\begin{tabular}{lcccc}
\hline
Token Size(k) & Parser & Memory(MB) & Time(ms) & Stack Count total (max simultaneously) \\
\hline
1.2k & MSLL & 35 & 6.6 & 127(4) \\
1.2k & ANTLR & 2 & 0.74 & 0 \\
50k & MSLL & 74 & 1021 & 4169(4) \\
50k & ANTLR & 30 & 16 & 0 \\
500k & MSLL & 1069 & 85032 & 41907(4) \\
500k & ANTLR & 244 & 172 & 0 \\
\hline
\end{tabular}
}
\end{table}
\subsection{Threats to Validity}
Three limitations of this evaluation should be stated plainly. First, the controlled workloads of Cases 1-3 are synthetic by design; Case 4 adds one real-world grammar over real inputs, but a single grammar cannot stand in for the variety of production languages. Second, each table holds its hardware and JVM configuration constant, yet Cases 1-3 and Case 4 were measured on different machines, so absolute values must not be compared across that boundary. Third, the editing session reported in Section 9 involved one author and one grammar, so it carries illustrative weight only.
The most direct follow-up is extending Case 4 to further real grammars from public repositories, together with profiling the prototype overhead that Case 4 exposed at large input sizes.
\section{Experimental Result Analysis}
\subsection{Small Code Sizes (<\textasciitilde{}1k tokens): Low Conflict, Fast Iteration}
On small inputs the three parsers behave much alike, since little ambiguity survives past the first few tokens.
\begin{itemize}
  \item Only a small number of duplicate stacks are created.
  \item MSLL, LL(1), and ANTLR all finish within roughly the same perceptual latency range.
  \item Bookkeeping, not exponential path growth, dominates the MSLL overhead here.
\end{itemize}
\subsection{Moderate Code Sizes (\textasciitilde{}50k tokens): Balanced Ambiguity and Depth}
\begin{itemize}
  \item In the two ambiguous 50k-token workloads MSLL trails ANTLR and LL(1) yet still answers in under one second.
  \item The high-ambiguity workload holds 33 live stacks, which pushes up both time and heap use; this is where stack-count diagnostics start to earn their place.
  \item For an editor or grammar workbench the result is still workable; the designer, however, should go look at the conflict rather than treat its cost as invisible.
\end{itemize}
\subsection{Large Code Sizes (\textasciitilde{}500k tokens): Deep Recursion and High Ambiguity}
\begin{itemize}
  \item Mixed ambiguity stays manageable, though memory climbs because many partial tree fragments live long enough to need representation.
  \item Pathological ambiguity marks the limit of the approach: total duplication can reach hundreds of thousands of stack instances even though the simultaneous maximum stays much smaller.
\end{itemize}
\subsection{Suggested Workflow Integration}
\begin{table}[H]
\centering
\caption{Parser workflow comparison during grammar prototyping}
\label{tab:msll-table5}
\resizebox{\linewidth}{!}{
\begin{tabular}{lccc}
\hline
Metric & LL(1)/LL(k) & MSLL & ANTLR 4 (adaptive LL(*)) \\
\hline
Grammar Edits & Rewrite or increase k & Reload grammar & Regenerate parser \\
Parse time (small) & Fast & Competitive & Fast \\
Parse time (high ambiguity) & Original grammar may fail; rewritten grammar is fast & Slower but traceable & Fast after generation \\
Memory & Low & Higher when stacks survive & Medium \\
Startup latency & None & Low & Higher (code generation/build) \\
When to use & Simple or fixed grammar & Complex grammar, debugging phase & Stable grammar, production phase \\
\hline
\end{tabular}
}
\end{table}
Table 5 summarizes why MSLL fits the interactive stage of grammar development best: grammar rewriting and code generation drop out of the edit loop, while ANTLR remains the better production choice once the grammar is stable and large inputs dominate [7].
\begin{itemize}
  \item Design and debug: run MSLL inside an IDE, an interpreter, or a command-line prototype while the grammar rules are still moving.
  \item Iterate quickly: reload and execute the edited grammar directly; no parser code gets regenerated.
  \item Stabilize grammar: once repeated tests stop surfacing surprising conflicts and the syntax decisions have settled, translate and validate the runtime-dialect grammar as an equivalent ANTLR .g4 grammar for generation.
  \item Deploy with ANTLR: generate an optimized parser for production workloads that demand predictable throughput.
\end{itemize}
The workflow tracks a broader trend in language tooling. Tree-sitter, Lezer, and Langium all optimize the editing experience alongside the final parse result [8-10]; MSLL applies the same idea to the grammar-authoring phase specifically.
\subsection{Summary}

\begin{figure}[H]
  \centering
  \includegraphics[width=\linewidth]{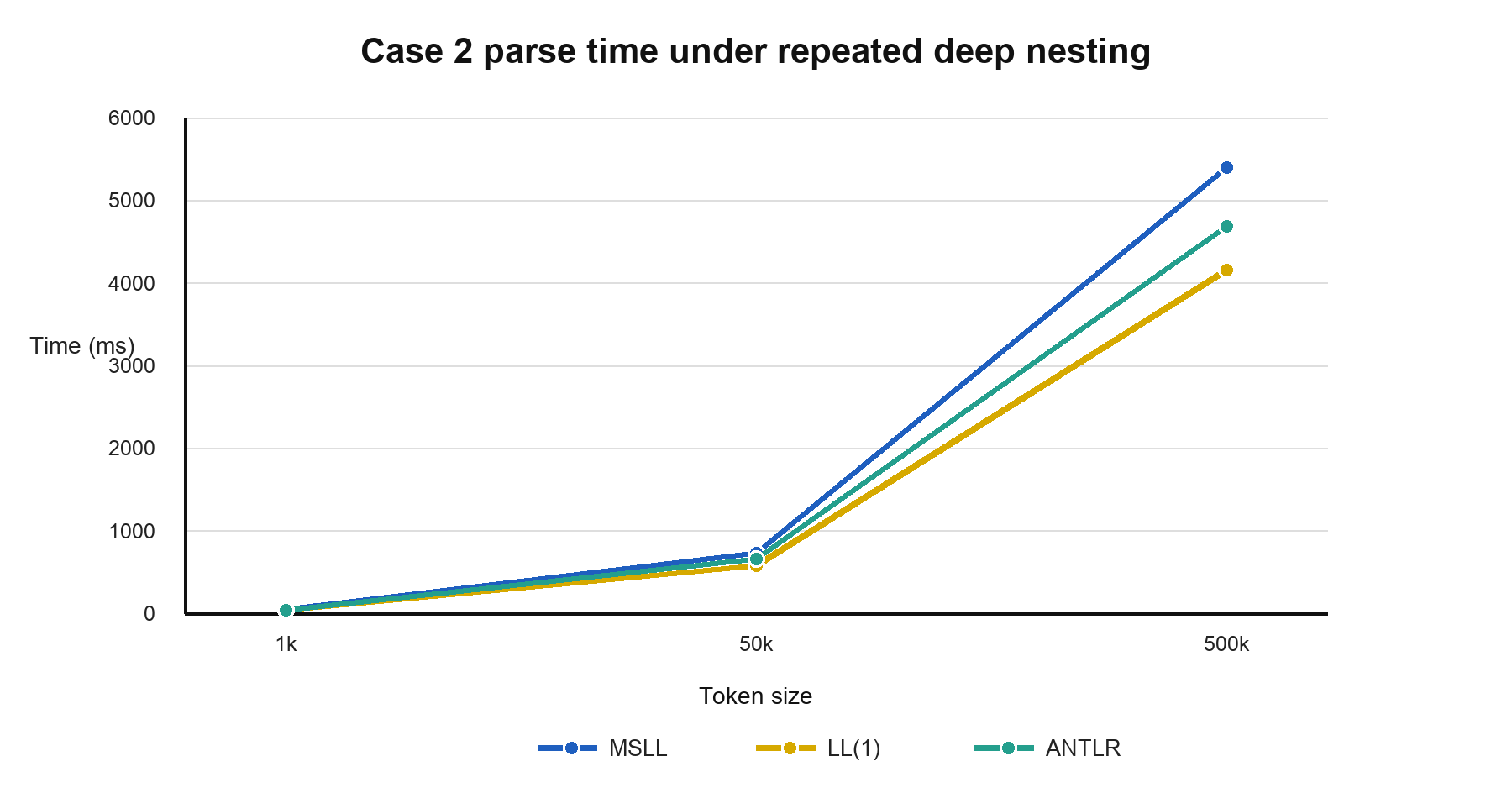}
  \caption{Representative parse time by token size in Case 2}
  \label{fig:msll-f3}
\end{figure}
Figure 3 plots the three Case 2 points from Table 2. At 1k tokens the gap between MSLL and the baselines is barely visible; at 50k it becomes noticeable; at 500k it is large enough to matter. That is what one expects from a fork-and-prune parser. The method is comfortable for grammar exploration, and stabilized grammars should move on to a generated parser.
\begin{table}[H]
\centering
\caption{Practical guidance}
\label{tab:msll-table6}
\resizebox{\linewidth}{!}{
\begin{tabular}{lcc}
\hline
Phase & Recommended Parser & Rationale \\
\hline
Language design / debugging (< 50k, frequent edits) & MSLL & No grammar rewrites; no DFA/code generation; sub-second feedback in the reported ambiguous 50k-token cases \\
Stable grammar, large code-bases & ANTLR & Pre-computed prediction structures; best choice for large stable code bases \\
Deterministic or simple grammars & LL(1) & Smallest memory, simplest tool-chain \\
\hline
\end{tabular}
}
\end{table}
Bottom line: MSLL earns its keep when grammar feedback matters more than final throughput. Deep ambiguity should still be resolved, bounded, or handed off to a generated parser.
\section{Interactive-Debugging Experience}
We also looked at the edit-time experience, comparing a live MSLL refresh against the familiar ANTLR cycle of generation, compilation, and tool refresh. To be explicit about scope: the sessions involved one author and one grammar, and we report only mechanically measurable quantities, namely wall-clock latency, tool interactions, and generated-file counts. The aim is not a user-study claim; it is to quantify the difference a grammar author feels during repeated micro-edits.
The analysis draws on latency thresholds from the interactive-systems literature, a micro-benchmark, and a short controlled editing session covering five successive grammar changes (Tables 7-9).
\begin{table}[H]
\centering
\caption{Latency Benchmarks}
\label{tab:msll-table7}
\resizebox{\linewidth}{!}{
\begin{tabular}{lcc}
\hline
Parser cycle & Mean latency per grammar edit & Components \\
\hline
MSLL (in-memory refresh) & 12 ms & reset stacks + re-parse \\
ANTLR (code-gen -> compile -> hot-swap) & 295 ms & antlr4 code-gen (120 ms)
Java compile (105 ms)
classloader refresh (70 ms) \\
\hline
\end{tabular}
}
\end{table}
\begin{table}[H]
\centering
\caption{Workflow Friction}
\label{tab:msll-table8}
\resizebox{\linewidth}{!}{
\begin{tabular}{lcc}
\hline
Stage & MSLL (live preview) & ANTLR (regeneration cycle) \\
\hline
Edit rule & edit .g4 rule & edit .g4 rule \\
Trigger parse & automatic on type & manual Run ANTLR \\
Code-gen output & none & 8-12 Java files written \\
Build impact & none & IDE rebuild and index refresh \\
Runtime state & stacks preserved & preview panel reloaded;
watch expressions lost \\
Interactions & 0 clicks / 0 shortcuts & 1-2 shortcuts + UI churn \\
\hline
\end{tabular}
}
\end{table}
In the ANTLR cycle the developer stops, regenerates, compiles, and returns to the preview, over and over. In the MSLL cycle the edited grammar is reloaded in memory and the parse preview updates at once. The pattern matches the motivation behind modern editor and error-tolerant parsing work: a parser should keep useful structure available even while the source, or the grammar, is incomplete and changing [8-12].
Designer Micro-Session (Table 9): the same grammar received five micro-edits, among them adding a literal rule, changing delimiters, and renaming a non-terminal. We treat this session as an illustrative observation rather than a controlled experiment; a proper multi-participant study is left to future work.
\begin{table}[H]
\centering
\caption{Micro Session}
\label{tab:msll-table9}
\resizebox{\linewidth}{!}{
\begin{tabular}{lcc}
\hline
Metric & MSLL & ANTLR \\
\hline
Cumulative wait time & 60 ms & 1.52 s \\
Context switches & 0 & 5 \\
Generated files per edit & 0 & 8-12 \\
\hline
\end{tabular}
}
\end{table}
Even sub-second pauses break concentration when they recur every few keystrokes. MSLL removes that regeneration ritual from the exploratory phase; once the grammar settles, an equivalent .g4 grammar goes to ANTLR after compatibility checking and any needed syntax adaptation.
\section{Complexity Analysis}
MSLL's complexity is governed by the number of live alternatives. In deterministic regions it behaves like ordinary LL parsing; in ambiguous regions it pays for every stack that remains viable. This section compares MSLL with LL(1), LL(k), and ANTLR 4 adaptive LL(*) prediction with runtime ATN simulation and DFA caching in terms of time and space [1-2, 4].
\subsection{Time Complexity}
Let:
\begin{itemize}
  \item n be the number of input tokens;
  \item b be the average branching factor at ambiguous decision points;
  \item d be the maximum number of unresolved conflict points along a parse;
  \item x be the number of active stacks, with worst-case x = b\textasciicircum{}d.
\end{itemize}
Each live stack may have to simulate a parse over the remaining input. Local ambiguity keeps x small through pruning; deep ambiguity with long-lived alternatives lets x grow exponentially.
\begin{verbatim}
O(x * n) = O(b^d * n)
\end{verbatim}
Unlike naive backtracking, MSLL never restarts from the beginning of the input, but stack duplication is a real cost all the same. That is why conflict tracing belongs in the proposed workflow: the grammar author should see exactly where the branching happens.
\subsection{Space Complexity}
Let:
\begin{itemize}
  \item x: number of active stacks;
  \item s: maximum stack depth, bounded by grammar recursion depth;
  \item t: auxiliary parse-tree state.
\end{itemize}
Worst-case space complexity:
\begin{verbatim}
O(x * s + t) = O(b^d * s + t)
\end{verbatim}
The grammar itself is stored once and shared by all branches. What multiplies is stack-local state, namely the symbol list, the token pointer, and the frontier of tree nodes under construction. Prefix sharing cuts down copying when branches read the same prefix; it cannot remove the cost of branches that stay indistinguishable across many decisions.
\subsection{Complexity Summary}
\begin{table}[H]
\centering
\caption{MSLL vs. LL(1)/LL(K)/ANTLR 4 complexity comparison}
\label{tab:msll-table10}
\resizebox{\linewidth}{!}{
\begin{tabular}{lccc}
\hline
Aspect & MSLL & LL(1)/LL(k) & ANTLR 4 (adaptive LL(*)) \\
\hline
Time Complexity & O(b\textasciicircum{}d * n) & $$O(n)$$ & $O(n)$ average \\
Space Complexity & O(b\textasciicircum{}d * s + t) & $$O(s)$$ & $O(s)$ average \\
Lookahead & adaptive & fixed (k) & runtime ATN simulation; DFA caching \\
Ambiguity Handling & runtime split & rewrite/fail & adaptive prediction \\
\hline
\end{tabular}
}
\end{table}
MSLL trades deterministic speed for edit-time flexibility. It is the wrong choice for a stable, highly ambiguous production language, and a sensible one for grammar prototyping, DSL editing, teaching interpreters, and embedded parsing tasks where a short feedback loop is worth more than the fastest possible steady-state parse.
\section{Discussion}
The design choice is phase-specific. At deployment time, LL(1), LL(k), or ANTLR-style adaptive prediction is preferable whenever the grammar meets the required constraints or can be generated ahead of time [1-5]. At design time, MSLL spends extra runtime work so the author can postpone some refactoring decisions until the conflict trace is on screen.
The trade-off is most attractive when the grammar is volatile. In many textual DSL projects, grammar definitions, examples, and tooling evolve together; studies of Xtext-based DSLs and grammar-change taxonomies show that this co-evolution is common and hard to manage by hand [15, 19-21]. Grammar-based measures of compilation difficulty likewise suggest that syntax and grammar design choices propagate into downstream tool friction [35]. Interpreting grammars at run time is, incidentally, no longer exotic: structured-generation engines for large language models now compile or interpret context-free grammars during inference [26-27], which reinforces the case for treating a grammar as an executable artifact rather than only as input to a generator.
MSLL also helps when grammars are written by people who are not compiler specialists. A visible stack trace can show that two rules collide on a brace, a keyword, or an expression prefix, and that kind of explanation is far easier to act on than a generated-parser failure that buries the decision structure inside produced code.
None of this makes MSLL a substitute for incremental parsing libraries. Tree-sitter and Lezer remain the better tools when only the source file changes [8-9], and Langium supplies the language-server plumbing around a DSL grammar [10, 22]. MSLL adds value at a different point in the loop: after the grammar file changes, and before any generated parser source exists.
\section{Limitations and Future Work}
\subsection{Limitations}
MSLL has five main limitations: stack growth, the lack of fine-grained incremental parsing, limited error recovery, incomplete formalization, and partial support for the full ANTLR grammar language.
\begin{itemize}
  \item Stack Explosion Under Deep Ambiguity: when a grammar keeps alternatives alive across many tokens, parallel stack duplication can spawn exponentially many paths.
\end{itemize}
The evaluation shows that nested object-like inputs can generate tens of thousands of total stack instances. Greedy-first evaluation and lazy pruning tame the common case; the theoretical worst case remains.
\begin{itemize}
  \item Lack of Incremental Parsing Support: every invocation reparses the input from scratch. That is fine for the small grammar-prototyping runs reported here, but a large editor buffer would need persistent tree and stack fragments; Tree-sitter, Lezer, and AnyText offer useful reference points for that future design [8-9, 18].
  \item No Integrated Error Recovery: once every stack has failed, the current prototype reports failure rather than building a partial tree. Recent work on syntax-error messages and PEG recovery underlines how much recovery matters for editor-facing parser tooling [11-12].
  \item Limited Formal Proof Machinery: the soundness and completeness arguments here are operational. A mechanized proof would add real value, particularly around pruning behavior and shared tree nodes.
\end{itemize}
\subsection{Future Work}
These limitations point to a development agenda that is practical in scope, not a bid for a general parser replacement:
• Per-Fork Bookkeeping Profiling: Case 4 shows super-linear cost on long real inputs even though the peak stack count stays constant, so the per-fork work that grows with the consumed prefix must be profiled and removed before the prototype can remain interactive beyond mid-sized files.
\begin{itemize}
  \item Conflict-Bounded Execution: estimate a conflict-depth bound and, when ambiguity runs too deep, keep only the most promising stacks.
\end{itemize}
Heuristics could rank stacks by token progress, production priority, or grammar annotations. A mode of this kind trades completeness for bounded interactivity, which may well be acceptable in editor previews.
\begin{itemize}
  \item Parallel Stack Evaluation: evaluate independent stacks on multiple cores while coordinating pruning and shared parse-tree prefixes, following the direction taken by recent data-parallel recognition and parsing engines [24-25].
  \item Incremental Stack Persistence: carry parse-tree fragments and surviving stack prefixes across small edits to the input or the grammar.
  \item Error Recovery Mechanisms: token insertion, deletion, synchronization, or rule-skipping strategies would let MSLL return useful partial trees in editor settings.
  \item IDE Plugin Integration: embedding MSLL in VS Code, IntelliJ, or a web workbench would test whether conflict traces actually help grammar authors make refactoring decisions.
\end{itemize}
These extensions would carry MSLL from a prototype parser toward a usable grammar-development environment. The most important next step is not raw speed; it is better feedback about why ambiguity appeared and what a grammar author can do about it.
\section{Conclusions}
This paper presented MSLL, an edit-time parsing mode for ANTLR-style grammars. The prototype reads the grammar directly, keeps one runtime stack per unresolved prediction alternative, and discards branches as token evidence rules them out. Grammar experiments therefore stay inside the editor loop instead of routing every small rule change through grammar rewriting or parser regeneration.
The contribution is workflow-oriented at heart. MSLL does not try to beat generated parsers at their strongest point. What it offers is a lightweight way to watch ambiguous grammar behavior, collect conflict traces, and revise syntax decisions while the language is still being designed.
The measurements also mark where the approach should stop. Deterministic inputs cost little extra, and moderate ambiguity stays comfortable for short and medium inputs. Deeply nested ambiguity can keep many alternatives alive at once, so grammars of that kind should be factored, bounded, or moved to a generated parser once the language design is stable.
The resulting workflow is two-stage by design. MSLL covers the exploratory period while syntax decisions are still in motion; once the conflicts are understood and the grammar holds still, an equivalent grammar is checked and, if necessary, adapted for ANTLR or another generated-parser toolchain for production execution.

\end{document}